\documentclass[runningheads]{llncs}
\usepackage{graphicx}
\usepackage{booktabs}
\usepackage{subfig}
\usepackage{caption}

\usepackage{tikz}
\usepackage{comment}
\usepackage{amsmath,amssymb} 
\usepackage{color}
\usepackage{orcidlink}
\usepackage[accsupp]{axessibility}  

\begin{document}
\pagestyle{headings}
\mainmatter
\def\ECCVSubNumber{346}  

\title{XCAT - Lightweight Quantized Single Image Super-Resolution Using Heterogeneous Group Convolutions and Cross Concatenation} 

\titlerunning{XCAT}
%
\author{Mustafa Ayazoglu \and
Bahri Batuhan Bilecen\index{Bilecen, Bahri Batuhan}}
\authorrunning{Ayazoglu et al.}
%
\institute{Aselsan Research, Ankara, Turkey
\\
\email{\{mayazoglu,batuhanb\}@aselsan.com.tr}}

\maketitle
\begin{abstract}
We propose a lightweight, single-image super-resolution mobile device network named XCAT, and introduce Heterogeneous Group Convolution Blocks with Cross Concatenations (HXBlock). The heterogeneous split of the input channels to the group convolution blocks reduces the number of operations, and cross concatenation allows for information flow between the intermediate input tensors of cascaded HXBlocks. Cross concatenations inside HXBlocks can also avoid using more expensive operations like 1x1 convolutions. To further prevent expensive tensor copy operations, XCAT utilizes non-trainable convolution kernels to apply upsampling operations. Designed with integer quantization in mind, XCAT also utilizes several techniques in training, like intensity-based data augmentation. Integer quantized XCAT operates in real-time on Mali-G71 MP2 GPU with 320ms, and on Synaptics Dolphin NPU with 30ms (NCHW) and 8.8ms (NHWC), suitable for real-time applications.
\end{abstract}

\section{Introduction}

\begin{figure}
\centering
\setlength{\tabcolsep}{0pt}

\begin{tabular}{cccc}

\begin{tabular}{@{}c@{}}
\captionsetup[subfigure]{position=bottom}
\subfloat[HR]{\includegraphics[width=0.47\linewidth]{"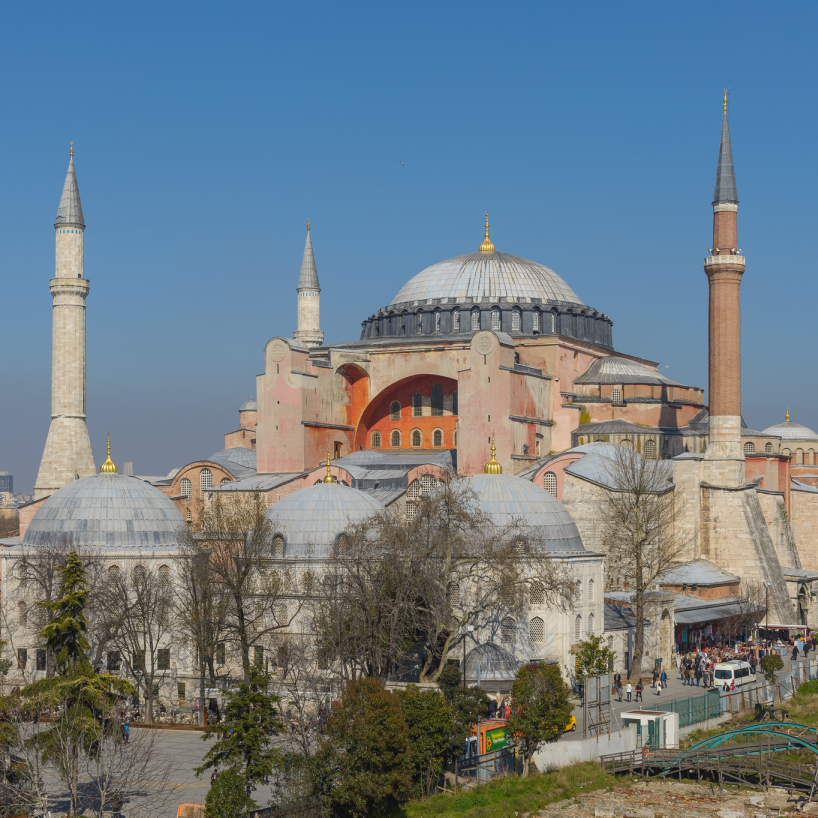"}} \\
\end{tabular} & 

\setlength{\tabcolsep}{0pt}
\begin{tabular}{@{}c@{}}
\captionsetup[subfigure]{position=bottom}
\subfloat[Portion \newline \centering of HR]{\includegraphics[width=0.17\linewidth]{"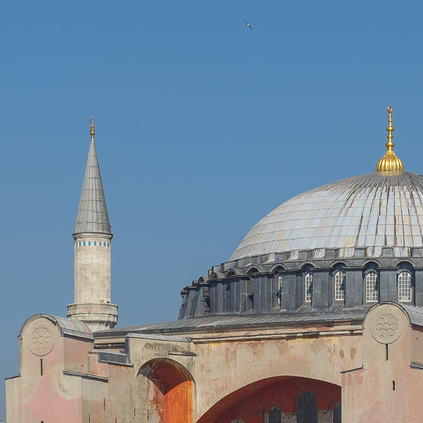"}}\\
\captionsetup[subfigure]{position=bottom}
\subfloat[\textbf{XCAT} \newline \centering (\textbf{320ms})]{\includegraphics[width=0.17\linewidth]{"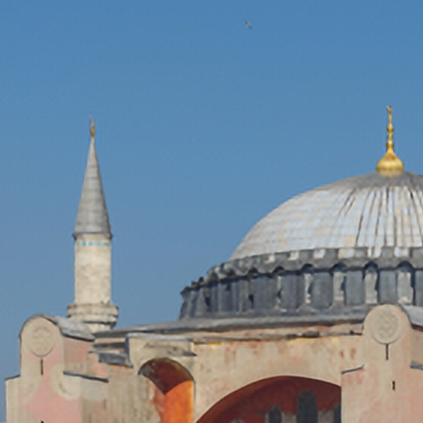"}} 
\end{tabular} & 

\begin{tabular}{@{}c@{}}
\captionsetup[subfigure]{position=bottom}
\subfloat[ESPCN\newline \centering \cite{Depth2Space}(363ms)]{\includegraphics[width=0.17\linewidth]{"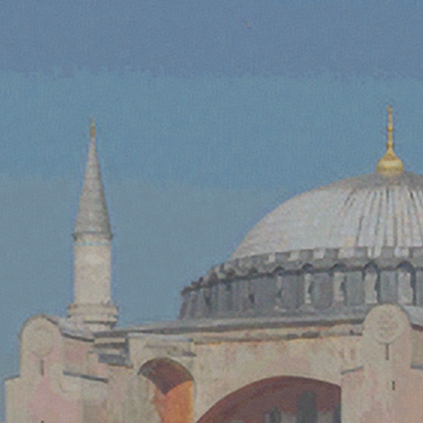"}}\\
\captionsetup[subfigure]{position=bottom}
\subfloat[ABPN\newline \centering \cite{ABPN}(600ms)]{\includegraphics[width=0.17\linewidth]{"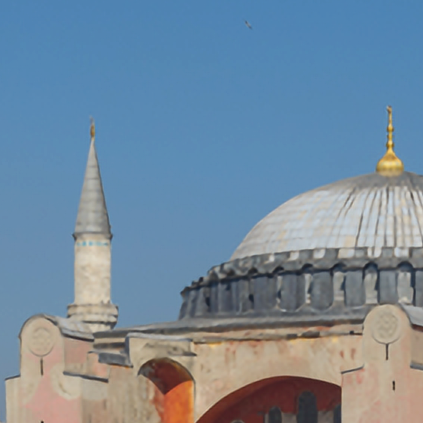"}}
\end{tabular} &

\begin{tabular}{@{}c@{}} 
\captionsetup[subfigure]{position=bottom}
\subfloat[FSRCNN\newline \centering \cite{FSRCNN}(485ms)]{\includegraphics[width=0.17\linewidth]{"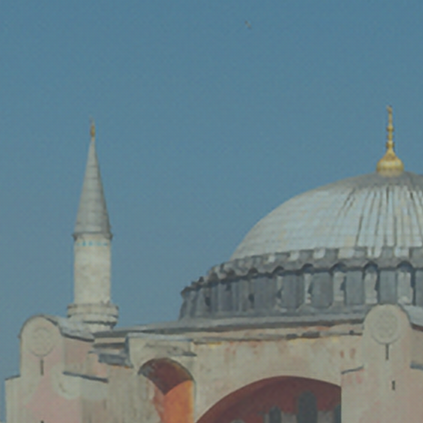"}}\\
\captionsetup[subfigure]{position=bottom}
\subfloat[XLSR\newline \centering \cite{XLSR}(370ms)]{\includegraphics[width=0.17\linewidth]{"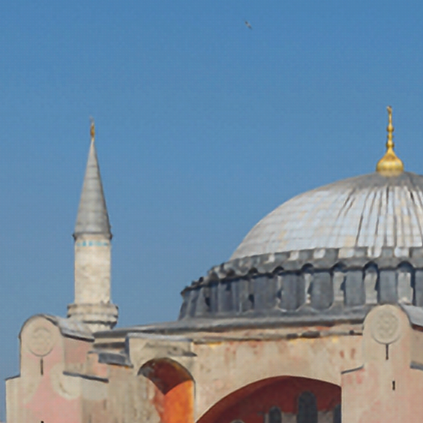"}}
\end{tabular}

\end{tabular}

\caption{Comparative results of UINT8 quantized models on DIV2K (Val image: 890). (d) and (f) yield both visually and numerically worse results than the rest. Visual results of (c), (e), and (g) are indistinguishable to the naked eye; however, XCAT runs the fastest}
\label{fig:hagia_sophia}
\end{figure}

Super-resolution (SR) is an extensively studied computer vision problem that aims to generate higher resolution (HR) image(s) given lower resolution (LR) image(s). In single image super-resolution (SISR), a single image; and in multi-image super-resolution (MISR), multiple images are utilized to generate a single HR image. In either case, image super-resolution is an ill-posed problem, since there is no unique solution. This ill-posed problem has been attempted to be solved via classical methods~\cite{SISR_Glasner} and deep-learning based methods~\cite{Depth2Space}\cite{SRCNN}; however, many new methods based on deep-learning are still being developed, most of which purely focus on data fidelity.

However, for the SR method to be practically applicable, the runtime is as important as the method's PSNR performance. Due to its practical importance, recent literature studies on SR focus on deployability, runtime, quantization, and efficiency, as well as PSNR of the method~\cite{XLSR}\cite{ABPN}\cite{IMDeception}\cite{MAI2021}\cite{EfficientSR2022}. Yet, achieving real-time performance with satisfactory visual quality during the quantization process further complicates the problem and careful network design is needed.

\begin{figure}
\begin{center}
\includegraphics[width=0.9\textwidth]{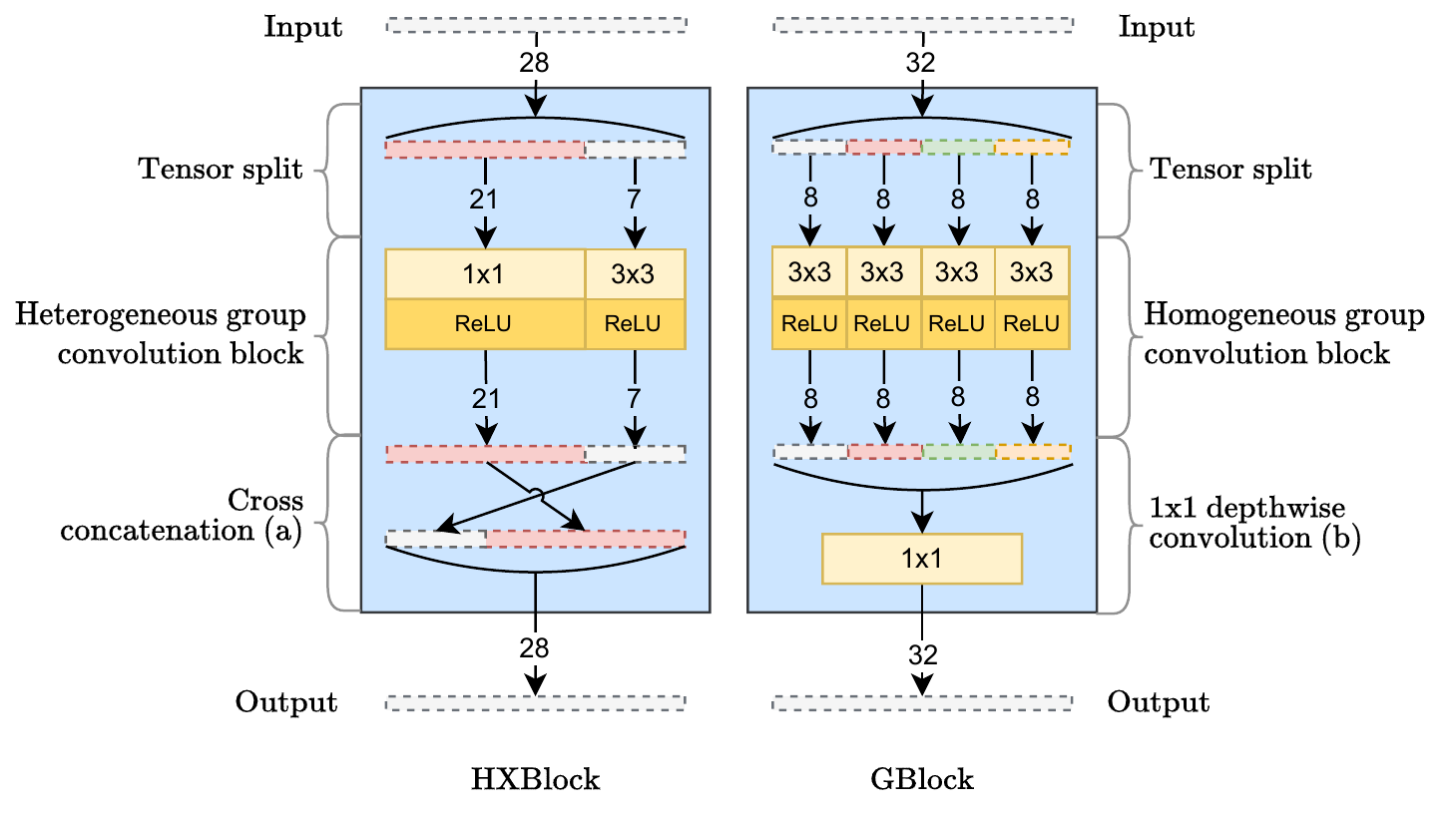}
\caption{Proposed HXBlock (left) vs. GBlock~\cite{XLSR} (right). Dashed rectangles represent tensors. Group convolutions inside HXBlock are \textbf{heterogeneous} compared to GBlock's. HXBlock uses \textbf{cross concatenation (a)} instead of depth wise 1x1 convolutions (b), which provides information flow through different convolutional kernels when HXBlocks are cascaded. For HXBlock, using (a) results in a significant runtime performance increase in return of a small PSNR drop compared to using (b)}
\label{fig:catvs1x1}
\end{center}
\end{figure}

In this study, we focus on the efficiency and mobile deployment, in the scope of \emph{Mobile AI \& AIM 2022 Real-Time Image Super-Resolution Challenge}~\cite{ignatov2022isr}. Our model, named \textbf{XCAT}, is a SISR network incorporating the proposed HXBlock and modifying both new and existing techniques from the literature for providing a quantization-aware, robust, real-time performance model, suitable for mobile devices.

Our work makes the following contributions:
\begin{enumerate}
    \item \textbf{HXBlocks}, which are heterogeneous grouped convolutions with cross concatenation layers for allowing information flow with almost no computational cost through different convolutional kernels of the group. Relevant studies done on HXBlocks have shown that they can be replaced with traditional group convolutions with little sacrifice from PSNR, but a significant gain in run time performance.
    \item A method for nearest neighborhood up sampling method with fixed 2D convolutional kernels to replace expensive tensor copy operations on mobile devices, which makes the model robust to quantization.
    \item An efficient, mobile device friendly, single image super-resolution network named \textbf{XCAT}.
\end{enumerate}

\section{Related Works}

\textbf{DNN-based single image super-resolution.} First deep-learning-based SISR algorithm was proposed by Dong et. al. as SRCNN~\cite{SRCNN}. Later, as a speed improvement on SRCNN, FSRCNN~\cite{FSRCNN} was developed; which introduced a deconvolutional layer at the end of the network, replaced ReLU with a PReLU activation layer, and reformulated SRCNN by adopting smaller filter sizes but more mapping layers. Shi et. al.~\cite{Depth2Space} introduced a novel, efficient sub-pixel convolutional layer (also known as depth to space), which is actually widely used in many fast SR networks right now. VDSR~\cite{VDSR}, EDSR~\cite{EDSR}, and WDSR~\cite{WDSR} continued the development of deep-learning-based SR by increasing the number of parameters, in exchange for accuracy with speed.

With the recent developments in computer vision and deep learning, concepts like attention mechanism~\cite{SISR_Holistic}, generative adversarial networks~\cite{SRGAN}\cite{ESRGAN}, recursive \& residual networks~\cite{DRCN}\cite{DRRN}\cite{CARN}, and distillation layers~\cite{IDN}\cite{IMDN}\cite{IMDeception}\cite{SwinTransformer} also started to take part inside SR network architectures. GANs and networks with attention mechanism mostly generate a high-quality SR image by sacrificing speed, whereas RNNs and distilling networks try to decrease the computational load.

\textbf{Group convolutions.} Group convolutions consist of groups of multiple convolutional kernels placed within the same layer. The motivation behind group convolutions emerged with AlexNet~\cite{AlexNet}, desiring to distribute the model over multiple GPUs to overcome hardware limitations. Later on, besides the increase in speed in AlexNet, group convolutions are also observed to improve classification accuracy when groups are accompanied by skipped connections with ResNetX~\cite{ResNetX}. ShuffleNet~\cite{ShuffleNet} introduced shuffling the intermediate tensors between group convolution blocks to increase feature extraction. DeepRoots~\cite{DeepRoots} and more recent studies use different convolutional kernels inside groups, such as 1x1 depth wise convolutions~\cite{GrPointwiseConv} and dilated convolutions~\cite{FirstDilatedConv}\cite{DSCR}. In addition, unitary~\cite{UnitaryGroup} and interleaved~\cite{InterleavedGroup} group convolutions also offer different perspectives on how to extract various features from input images. Usage of group convolutions due to their efficiency on super-resolution problems is also present~\cite{XLSR}\cite{IMDeception}\cite{EffSRLearnedGroupConv}.

\textbf{Model Optimization.} Hardware limitations and specifications may require the model to be optimized via different techniques, such as quantization, pruning, clustering, network architecture search (NAS), and many more. \textbf{Quantization} refers to converting floating point values to integers, hence decreasing memory usage and computational cost when re-accessing and/or updating the mentioned values, at a cost of decreasing the precision. Quantization is particularly useful on neural network models since it can decrease inference times without sacrificing much inference accuracy if done correctly~\cite{IntOnlyQuant}. Models also can be quantized after quantization-aware training in floating point precision~\cite{ImprovingNNonCPUs}\cite{QuantizationWhitePaper}, as well as training the network directly with low precision multiplications~\cite{NNwithLowPrecisionMult}. Removing layers from a model having a minor effect on inference is called \textbf{pruning}~\cite{PruningSurvey}\cite{ChannelPruning}, and \textbf{clustering} is the method of decreasing the number of unique weights by grouping weights and assigning the centroid values for each group. All of these methods try to decrease processor utilization or memory usage, or both.
Besides optimizing an existing network structure, finding the most possible optimal network structure in search space is also a study area, known as network architecture search~\cite{NAS}.

\section{Method}
In this section, XCAT is defined by its overall architecture and its components. Details about the training techniques and the quantization procedure are explained thoroughly as well.

\subsection{XCAT's Architecture}

\begin{figure}
\begin{center}
\includegraphics[width=0.9\textwidth]{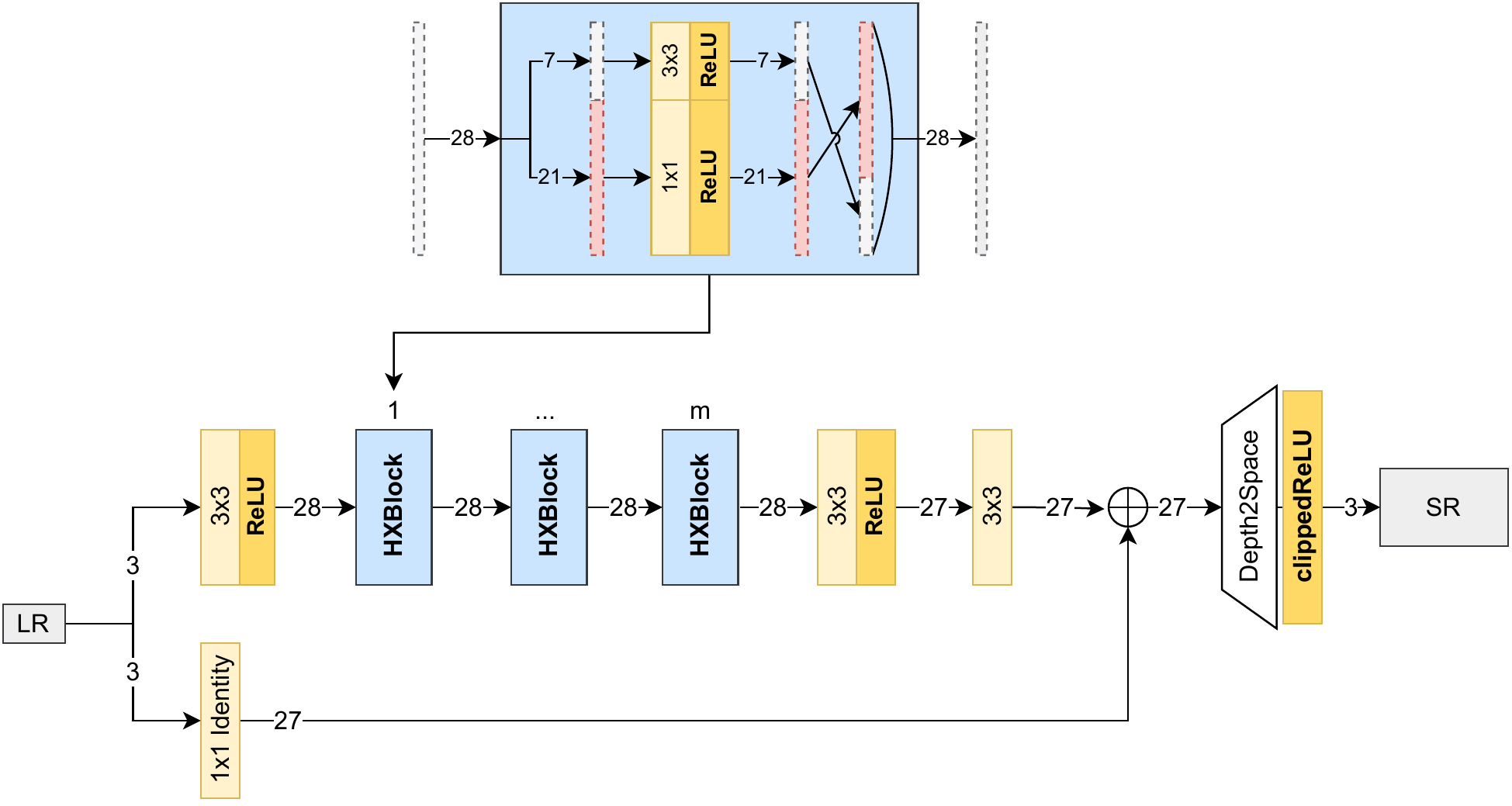}
\caption{Network structure of XCAT. Numbers on arrows denote channel numbers, and numbers inside blocks represent kernel sizes of convolutional blocks. Dashed blocks represent tensors, whereas non-dashed ones represent operators like convolution and normalization. Convolution with 1x1 identity kernel performs the upsampling method visualized in Fig.~\ref{fig:tensorcopy}. XCAT has m HXBlocks, where m=2 for this study}
\label{fig:overall}
\end{center}
\end{figure}

As seen from Fig.~\ref{fig:overall}, XCAT consists of 3 individual convolutional layers with trainable 3x3 kernels, a single 1x1 convolutional layer with a fixed "identity kernel" to simulate nearest neighborhood upsampling operation, m HXBlocks, a tensor addition layer, followed by a depth to space (D2S) layer and a clipped ReLU activation layer. Input and output of XCAT are LR and SR, respectively, where SR has x3 resolution of LR.

Each key component of XCAT will be detailed with their reasoning:

\textbf{Group convolutions with heterogeneous filter groups and varying kernels.} 
Group convolutions, which include multiple convolutional kernels per layer, are known to be able to extract and learn more varying features compared to a single kernel~\cite{GroupConv}. XCAT inherits this idea of group convolution blocks to replace single-layered convolutions in a repeated manner. However, as opposed to initial approaches~\cite{ResNetX}, convolutional layers inside the group convolution blocks in XCAT have \textbf{different layer dimensions} and \textbf{different kernel sizes} (Fig.~\ref{fig:overall}). This allows to pass the same source information between different convolutional layers and allows for less computationally demanding feature extraction. The input tensor is split into two parts in channel dimension, one processed by 1x1 and the other by 3x3 convolutional kernels. 1x1 convolution "blends" the point-wise information from previous HXBlocks and extracts inter-channel features, whereas 3x3 convolution considers in-channel correlation as well. In addition, a relevant study of Lee et al.'s~\cite{LogFilter} logarithmic filter groups in shallow CNNs shows the positive effect of dividing group convolution input tensors unevenly.

\textbf{Cross concatenation.} First group convolutions in AlexNet~\cite{AlexNet} ended with max pooling layers. However, group convolution designs such as DeepRoots~\cite{DeepRoots} started utilizing low-dimensional embeddings (like 1x1 convolutions) at the end of the groups, with the inspiration taken from Lin. et. al.~\cite{NetworkInNetwork} and Cogswell et. al.~\cite{Decorrelating}. This was done to decrease the computational cost and number of parameters without compromising accuracy. Later on, several efficiency-oriented SR networks like XLSR~\cite{XLSR} and IMDeception~\cite{IMDeception} also utilized group convolutions ending with 1x1 depth-wise convolutions.

In XCAT, instead of using 1x1 depth wise convolutions for increasing the spatial receptive field of each output of a group convolution block, the output tensor of each group convolution block is \textbf{cross concatenated}. The inspiration came from ShuffleNet~\cite{ShuffleNet} and Swin Transformer~\cite{SwinTransformer}; where channel shuffling and convolutional layers are inserted between group convolutions in the former, and window partitions are cyclic shifted to enable information flow between windows in the latter. Each cross concatenation in XCAT corresponds to a circular shift of one-fourth of the input tensor (Fig.~\ref{fig:overall}). This cyclic procedure allows the information to pass through from 1x1 and 3x3 convolutions inside XCAT's group convolution blocks, hence having more chance for feature extraction. 

It is worthy to note that ShuffleNet~\cite{ShuffleNet}'s channel shuffling is similar to the XCAT's; however, XCAT has a cross concatenation operation represented with cyclic shifts, whereas ShuffleNet has a shuffle operation dividing and reorganizing tensors into many small partitions. This reorganization operation is reflected onto the target device (Synaptics Dolphin NPU) as reshape and transpose operations, which take much longer to process compared to XCAT's simpler yet effective approach.

During the experiments, it is observed that replacing cross concatenation operations with 1x1 convolutions in XCAT increases the run time per frame, but does not increase the PSNR test score considerably, making it less practical for mobile networks.

\textbf{Depth to space (D2S) operation.} Shi et al.~\cite{Depth2Space}'s pixel shuffling (depth to space operator) is inserted at the end of the network, which aims to implement sub-pixel convolutions in an efficient manner and is proven to increase PSNR score in super-resolution problems in many studies.

\textbf{Nearest neighborhood upsampling with fixed kernel convolutions.} We observed that providing the low-resolution input image to D2S with accompanying feature tensors increases the robustness, as opposed to only providing the extracted feature tensors to D2S. With this motivation, XCAT also adds repeated input image tensors (where each channel of the input image is repeated 9 times) to feature tensors and provides them to D2S. From the perspective of D2S, this operation is equivalent to the nearest neighborhood upsampling.

\begin{figure}
\begin{center}
\includegraphics[width=0.6\textwidth]{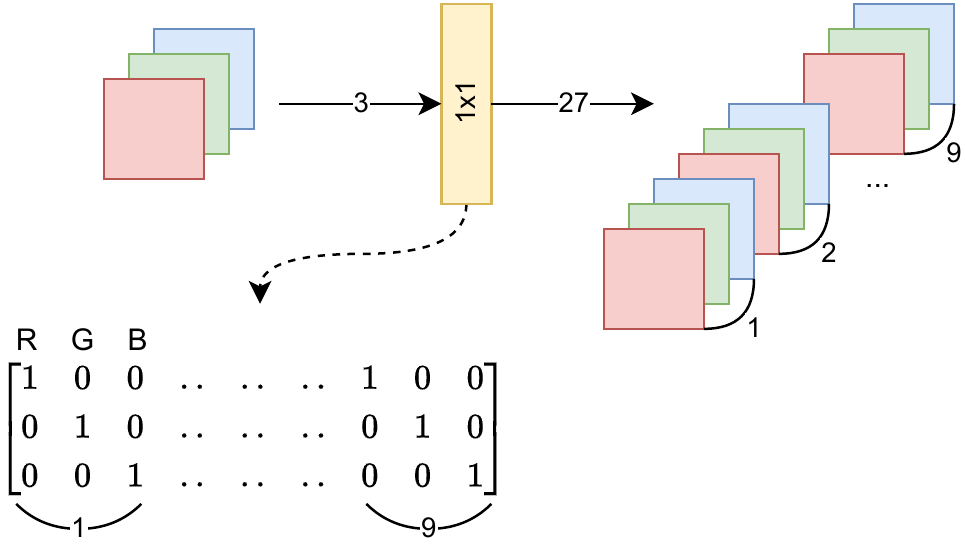}
\caption{Tensor copy operations done with convolutions. The identity kernel of the 1x1 convolution is set in such a way that it reproduces the input tensor of 3 channels 9 times, generating an output tensor of 27 channels}
\label{fig:tensorcopy}
\end{center}
\end{figure}

A relevant study done by Du et. al. named ABPN~\cite{ABPN} also utilizes the nearest neighborhood upsampling to be fed to the D2S block. However, it uses tensor copy operations while repeating and concatenating the input image in the upsampling process, which are indeed expensive for mobile devices. For a better alternative, a convolutional layer of 3 input channels and 27 output channels is used, with a 1x1 non-trainable kernel which is set to serve the same purpose as a tensor copy (Fig. \ref{fig:tensorcopy}). One point to note is that when this 1x1 kernel is set as trainable, it gets affected by the quantization process and yields lower visual quality results.

\subsection{Training and Quantization Details}

XCAT is trained in floating point precision and quantized afterward. However, it is trained and designed with quantization in mind, with several techniques to avoid PSNR decrease:

\textbf{Intensity-augmented training.} To minimize the PSNR difference while quantizing the FP32 model to its UINT8, intensity values of the training images are scaled with randomly chosen constants among (1, 0.7, 0.5). We have observed that this strategy helped with quantization and avoided signal degradation, as stated in~\cite{XLSR}.

\textbf{Clipped ReLU.} As proven and explained in~\cite{XLSR}, using clipped ReLU at the end of the network allows better quantization while keeping the performance in the real-time range.

\textbf{Representative dataset selection.} TensorFlow Lite requires a representative dataset while quantizing a floating point Keras or TensorFlow model. As a rule of thumb, this dataset consists of entire training images. However, it is observed that selecting a subset of all training images as the representative dataset affects the final PSNR test score of the quantized UINT8 model immensely. Hence, to find the most suitable representative dataset, a linear search is applied to all DIV2K training images, generating single image representative datasets. For each representative dataset (or rather an image), XCAT is quantized and PSNR test scores are measured. The highest scoring quantized XCAT model is chosen as the best.

\textbf{Training details.} XCAT is trained twice in floating point precision and then quantized. Training details are as follows:
\begin{itemize}
\item DIV2K dataset is used for the first training, and Flickr2K dataset is added alongside for the second training (fine tuning).
\item Intensity, rotation, random crop, and flip augmentations are used while setting up the dataset for both of the training. HR images are cropped to 96x96 patches.
\item XCAT is trained with 50 epochs and 16 batches. Each epoch contains 10000 mini-batches.
\item For the first training:
\begin{itemize}
\item Charbonnier loss is used, where $C(x) = \sqrt{(x^2 + {\epsilon}^2)}$ and $\epsilon=0.1$. Charbonnier loss is the smoother version of L1-loss having better convergence characteristics than L2-loss.
\item Adam optimizer is used with initial learning rate = 0.001, $\beta_{1}=0.9$, $\beta_{2}=0.999$, $\epsilon=1e^{-8}$.
\item Warm-up scheduler is used: Starting from the initial learning rate, at each epoch, the learning rate is increased up to $25e^{-4}$ until the 5th epoch. After the 5th epoch, the learning rate is linearly decreased at the end of each epoch, where at the last epoch it decreases to $e^{-4}$.
\end{itemize}
\item For the second training for fine-tuning:
\begin{itemize}
\item Mean square error is used as the loss.
\item Adam optimizer is used with an initial learning rate = 0.0001, and the same beta and epsilon parameters.
\item Warm-up scheduler is used again, but with the new initial learning rate, and the maximum learning rate of $12.5e^{-4}$ instead of  $25e^{-4}$.
\end{itemize}
\end{itemize}

\begin{table}
    \caption{Different XCAT-based models and their performance. Runtime is evaluated on Mali-G71 MP2 GPU via AI Benchmark 5~\cite{AIBenchmark}. Score is the metric function described in (\ref{eq:customscore}). Note that the Config column describes the differences among models. m is the number of HXBlocks. X/Y shows the splitting ratio, where X+Y is the total number of channels of HXBlocks. axa/bxb represents the convolutional kernels inside the group convolution blocks, where the tensor with dimension X passes through axa, and Y through bxb}
    \begin{center}
    \smallskip
    \begin{tabular}{ccccc}
        \toprule
        {\textbf{Model}} & {\textbf{Config}} &{\begin{tabular}{@{}c@{}}\textbf{PSNR} \\ \textbf{FP32/UINT8} \end{tabular}}  & {\begin{tabular}{@{}c@{}}\textbf{Runtime} \\ \textbf{(ms)} \end{tabular}} & {\textbf{Score}}\\
        \midrule
        \midrule
        \textbf{XCAT}& \scriptsize \begin{tabular}{@{}c@{}}m=2, 21/7, 1x1/3x3 \\ 2 stage training \end{tabular} & 29.88/29.81 & 320 & \textbf{240}\\
        \midrule
        A & \scriptsize \begin{tabular}{@{}c@{}}m=2, 21/7, 1x1/3x3 \\ 1 stage training \end{tabular} & 29.85/29.79 & 320 & 233\\
        \midrule
        B & \scriptsize \begin{tabular}{@{}c@{}}m=2, 21/7, 1x1/3x3 \\ Cross Cat $\rightarrow$ 1x1 Conv. \end{tabular} & 29.92/29.84 & 340 & 235\\
        \midrule
        C & \scriptsize \begin{tabular}{@{}c@{}}m=2, 21/7, 3x3/3x3 \\ Cross Cat $\rightarrow$ 1x1 Conv. \end{tabular} & \textbf{30.04/29.96} & 780& 121\\
        \midrule
        D & \scriptsize \begin{tabular}{@{}c@{}}m=2 \\ HXBlock $\rightarrow$ 3x3 Conv. \end{tabular} & 30.04/29.97 & 770& 124\\
        \midrule
        E & \scriptsize \begin{tabular}{@{}c@{}}m=4, 21/7, 1x1/3x3 \end{tabular} & 29.98/29.89 & 370 & 232\\
        \midrule
        F & \scriptsize \begin{tabular}{@{}c@{}}m=4, 21/7, 1x1/3x3 \\ Conv after last HXBlock: \\ 3x3 $\rightarrow$ 1x1 \end{tabular} & 29.82/29.75 & 300 & 236\\
        \midrule
        G & \scriptsize \begin{tabular}{@{}c@{}}m=4, 21/7, 1x1/3x3 \\ Conv after last HXBlock: \\ removed \end{tabular} & 29.81/29.72 & 290 & 234\\
        \midrule
        H & \scriptsize \begin{tabular}{@{}c@{}}m=4, 16/12, 1x1/3x3 \\ Conv after last HXBlock: \\ removed \end{tabular} & 29.87/29.76 & 300 & 238\\
        \midrule
        I & \scriptsize \begin{tabular}{@{}c@{}}m=4, 7/21, 1x1/3x3 \\ Conv after last HXBlock: \\ removed \end{tabular} & 30.03/29.88  & 520& 163\\
        \midrule
        J & \scriptsize \begin{tabular}{@{}c@{}}m=4, 7/21, 3x3/3x3 \\ Conv after last HXBlock: \\ removed \end{tabular} &  30.04/29.87 & 550& 152\\
        \midrule
        K & \scriptsize \begin{tabular}{@{}c@{}}m=3, 7/21, 1x1/3x3 \\ Conv after last HXBlock: \\ removed \end{tabular} & 29.95/29.81 & 430& 179\\
        \midrule
        L & \scriptsize \begin{tabular}{@{}c@{}}m=4, 16/4, 1x1/3x3 \\ Conv after last HXBlock: \\ removed \end{tabular} & 29.61/29.45 & \textbf{205}& 228\\
        \midrule
        M & \scriptsize \begin{tabular}{@{}c@{}}m=4, 16/4, 1x1/3x3 \\ Replaced Add with Concat \end{tabular} & 29.63/29.15 & 298& 103\\
        \bottomrule
        \bottomrule
    \end{tabular}
    \end{center}
    \label{tab:psnr}
\end{table}

\begin{table}
     \caption{PSNR test scores of XCAT and other algorithms with public datasets. All models are FP32. To be consistent with the rest of the algorithms; XLSR, ABPN, and XCAT's PSNR results are calculated using Luminance (Y) channel rather than RGB channels, except for DIV2K (*We performed our own training since the pre-trained FP32 model from the authors performed poorly, around 15dB for DIV2K(Val))}
    \begin{center}
    \begin{tabular}{cccccccc}
    \toprule
    \textbf{Dataset} & \textbf{Scale} & \textbf{Bicubic} & \textbf{FSRCNN} & \textbf{ESPCN} & \textbf{XLSR} & \textbf{ABPN}* & \begin{tabular}{@{}c@{}} \textbf{XCAT} \\ (proposed) \end{tabular}\\
    \midrule
    {Set5} & {x3} & {30.44} & {32.73} & {32.59} & {33.09} & {33.45} & {33.02}\\
    
    {Set14} & {x3} & {27.63} & {29.30} & {29.18} & {29.59} & {29.73} & {29.54}\\
    
    {B100} & {x3} & {27.13} & {28.26} & {28.18} & {28.45} & {28.56} & {28.42}\\
    
    {Urban100} & {x3} & {24.43} & {26.03} & {25.87} & {26.48} & {26.73} & {26.38}\\
    
    {Manga109} & {x3} & {26.87} & {30.21} & {29.70} & {31.13} & {31.47} & {31.12}\\
    
    {DIV2K(Val)} & {x3} & {28.82} & {29.67} & {29.54} & {30.10} & {30.10} & {29.88}\\
    
    \bottomrule
    \end{tabular}
    \end{center}
    \label{tab:FP32_PSNR_v2}
\end{table}

\begin{table}
     \caption{Effect of cross concatenation versus straight concatenation (no shuffling of the tensors while concatenating), the number of HXBlocks, and the tensor divisions. All models are based on XCAT. All parameter changes are mentioned on the table, and the rest are kept the same among all models. Runtime is evaluated on Mali-G71 MP2 GPU via AI Benchmark 5~\cite{AIBenchmark}. Score is the metric function described in (\ref{eq:customscore}). Split and kernel definitions are stated in Tab.~\ref{tab:psnr}}
    \begin{center}
    \begin{tabular}{cccccccc}
    \toprule
    \textbf{Split} & \textbf{Kernel} & \begin{tabular}{@{}c@{}} \textbf{m} \\ \textbf{(\# of HXBlocks)} \end{tabular}  & \begin{tabular}{@{}c@{}} \textbf{Cross} \\ \textbf{Concat} \end{tabular} & \begin{tabular}{@{}c@{}} \textbf{PSNR} \\ (FP32) \end{tabular} & \begin{tabular}{@{}c@{}} \textbf{PSNR} \\ (UINT8) \end{tabular} & \begin{tabular}{@{}c@{}} \textbf{Runtime} \\ (ms) \end{tabular} & \textbf{Score} \\
    \midrule
    {21/7} & {1x1/3x3} & {2} & {\checkmark} & {29.88} & {29.81} & {320} & {240} \\
    
    {21/7} & {1x1/3x3} & {2} & {$\times$} & {29.87} & {29.79} & {320} & {233}\\
    \midrule
    {21/7} & {1x1/3x3} & {4} & {\checkmark} & {29.98} & {29.89} & {370} & {232}\\
    
    {21/7} & {1x1/3x3} & {4} & {$\times$} & {29.96} & {29.87} & {370} & {232} \\
    \midrule
    {21/7} & {1x1/3x3} & {8} & {\checkmark} & {30.04} & {29.97} & {480} & {200}\\
    
    {21/7} & {1x1/3x3} & {8} & {$\times$} & {30.01} & {29.89} & {480} & {179} \\
    \midrule
    {21/7} & {1x1/3x3} & {12} & {\checkmark} & {30.07} & {29.97} & {580} & {165}\\
    
    {21/7} & {1x1/3x3} & {12} & {$\times$} & {30.04} & {29.72} & {580} & {117} \\
    \midrule

    {24/8} & {1x1/3x3} & {6} & {\checkmark} & {30.02} & {29.92} & {410} & {218} \\ 
    
    {24/8} & {1x1/3x3} & {6} & {$\times$} & {30.01} & {29.89} & {410} & {209} \\ 
    \midrule

    {56/8} & {1x1/3x3} & {4} & {\checkmark} & {30.08} & {30.03} & {620} & {168} \\ 
    
    {56/8} & {1x1/3x3} & {4} & {$\times$} & {30.04} & {29.99} & {620} & {159} \\
    
    \bottomrule
    \end{tabular}
    \end{center}
    \label{tab:XCrossMethods}
\end{table}

\section{Experimental Results}

\begin{figure}
\centering
\centering
\setlength{\tabcolsep}{0pt}
\begin{tabular}{cccc}
\begin{tabular}{@{}c@{}} 
\captionsetup[subfigure]{position=bottom}
\subfloat[HR]{\includegraphics[width=0.41\linewidth]{"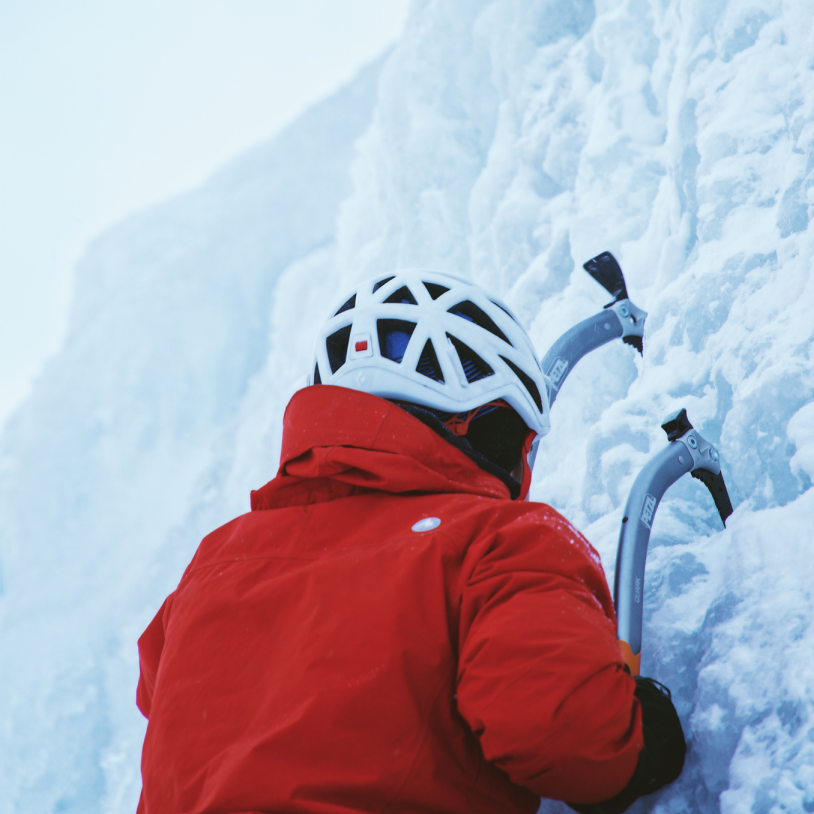"}} \\ 
\end{tabular} & 

\setlength{\tabcolsep}{0pt}
\begin{tabular}{@{}c@{}}
\captionsetup[subfigure]{position=bottom}
\subfloat[HR]{\includegraphics[width=0.17\linewidth]{"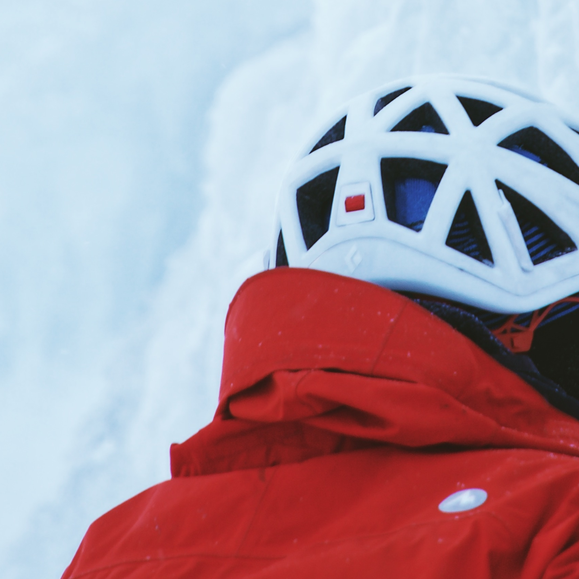"}}\\
\captionsetup[subfigure]{position=bottom}
\subfloat[\textbf{XCAT}]{\includegraphics[width=0.17\linewidth]{"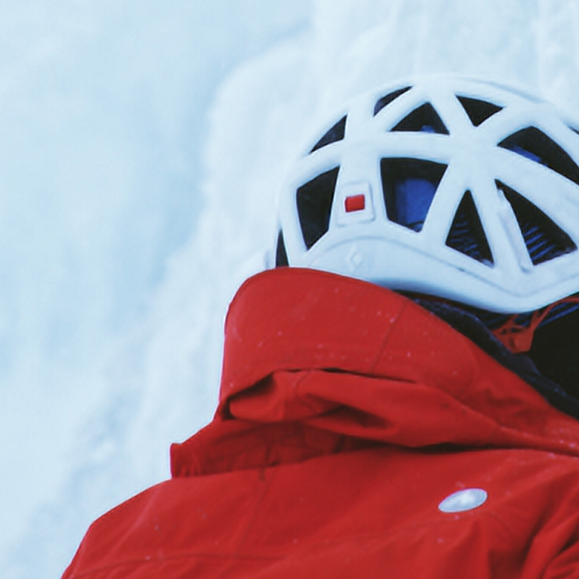"}} 
\end{tabular} & 
\begin{tabular}{@{}c@{}}
\captionsetup[subfigure]{position=bottom}
\subfloat[ESPCN]{\includegraphics[width=0.17\linewidth]{"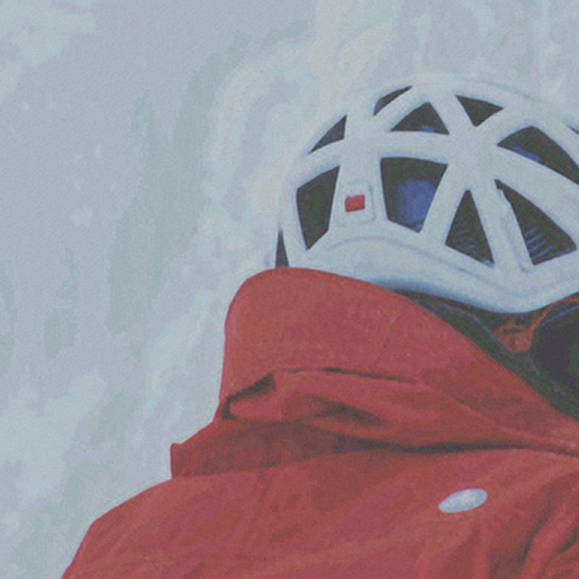"}}\\
\captionsetup[subfigure]{position=bottom}
\subfloat[ABPN]{\includegraphics[width=0.17\linewidth]{"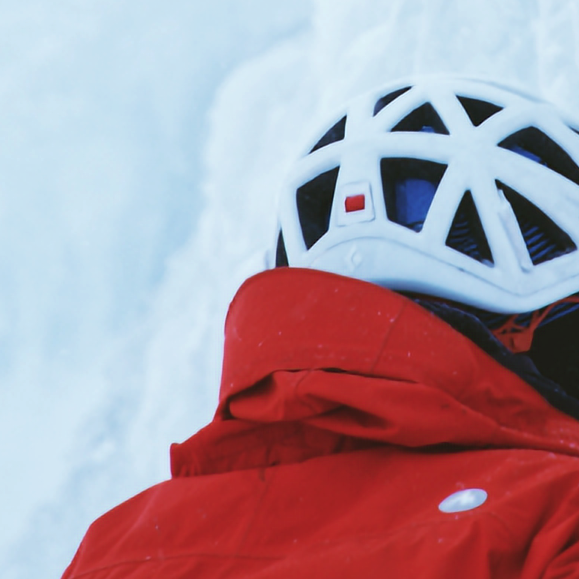"}}
\end{tabular} &
\begin{tabular}{@{}c@{}} 
\captionsetup[subfigure]{position=bottom}
\subfloat[FSRCNN]{\includegraphics[width=0.17\linewidth]{"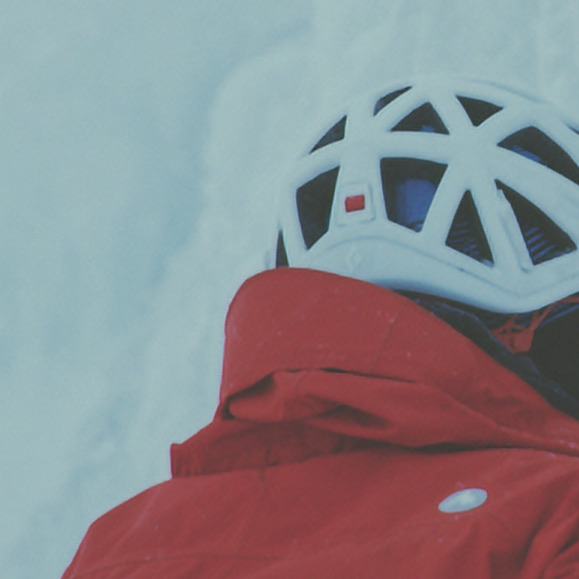"}}\\
\captionsetup[subfigure]{position=bottom}
\subfloat[XLSR]{\includegraphics[width=0.17\linewidth]{"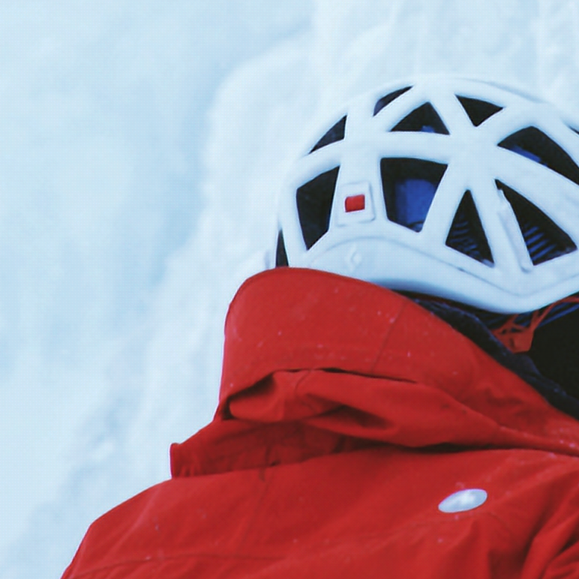"}}
\end{tabular}
\end{tabular}


\centering
\captionsetup[subfigure]{labelformat=empty}
\setlength{\tabcolsep}{0pt}

\begin{tabular}{cccc}
\begin{tabular}{@{}c@{}} 
\captionsetup[subfigure]{position=bottom}
\subfloat[(a) HR]{\includegraphics[width=0.41\linewidth]{"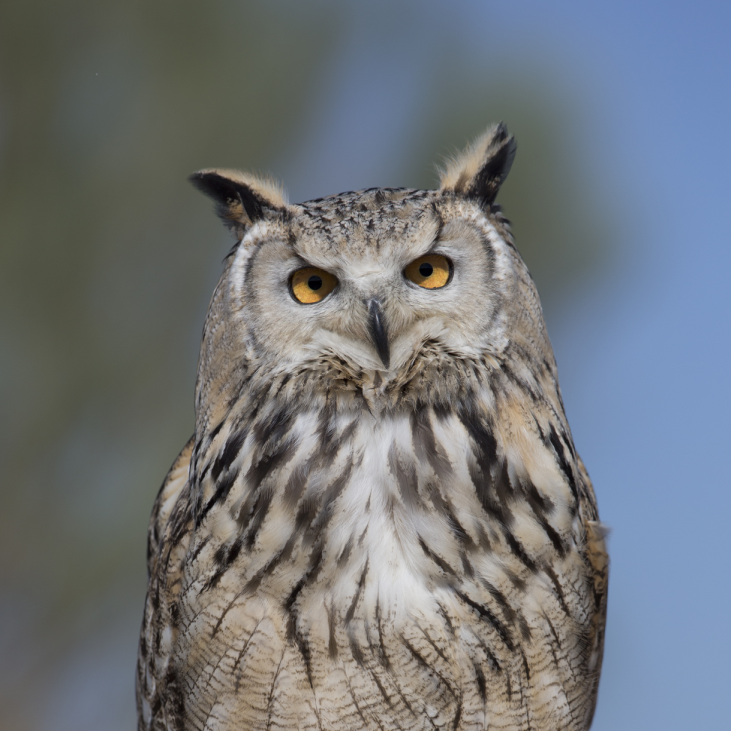"}} \\ 
\end{tabular} & 
\setlength{\tabcolsep}{0pt}

\begin{tabular}{@{}c@{}}
\captionsetup[subfigure]{position=bottom}
\subfloat[(b) HR]{\includegraphics[width=0.17\linewidth]{"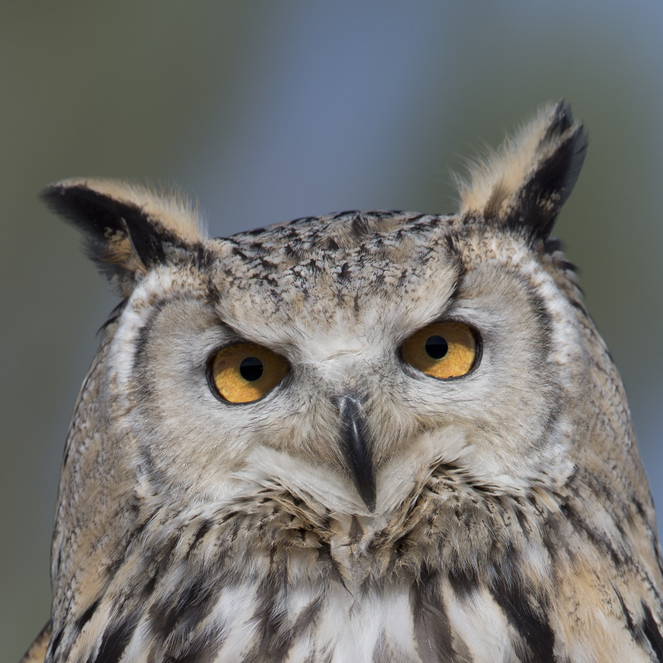"}}\\
\captionsetup[subfigure]{position=bottom}
\subfloat[(c) \textbf{XCAT}]{\includegraphics[width=0.17\linewidth]{"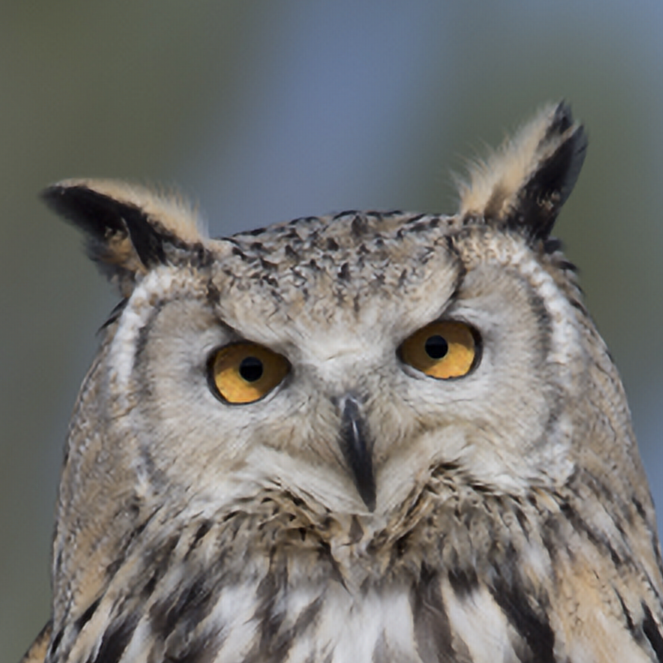"}} 
\end{tabular} & 
\begin{tabular}{@{}c@{}}
\captionsetup[subfigure]{position=bottom}
\subfloat[(d) ESPCN]{\includegraphics[width=0.17\linewidth]{"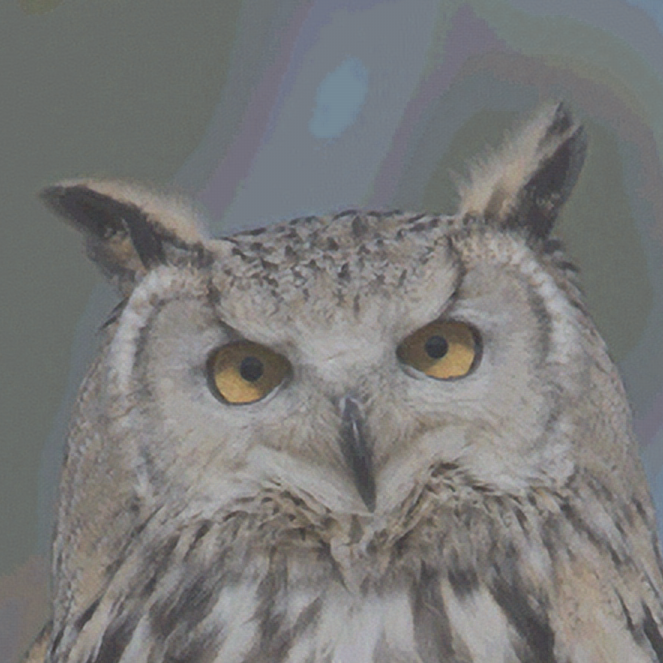"}}\\
\captionsetup[subfigure]{position=bottom}
\subfloat[(e) ABPN]{\includegraphics[width=0.17\linewidth]{"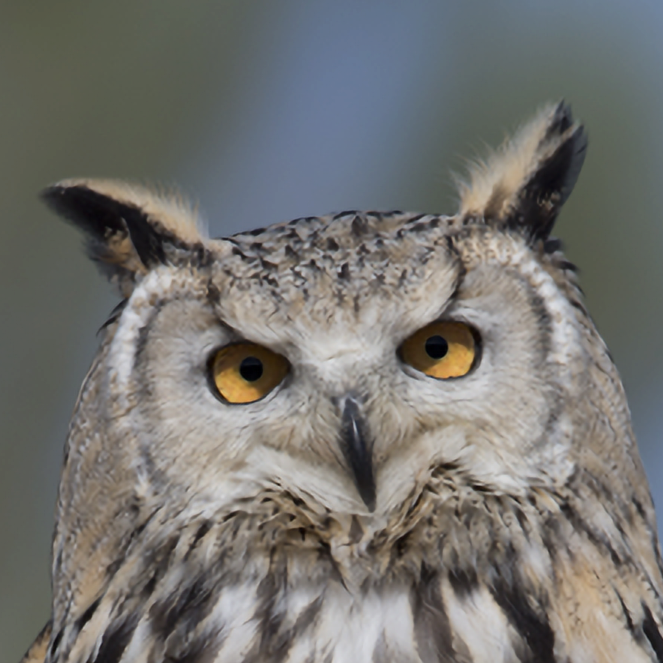"}}
\end{tabular} &
\begin{tabular}{@{}c@{}} 
\captionsetup[subfigure]{position=bottom}
\subfloat[(f) FSRCNN]{\includegraphics[width=0.17\linewidth]{"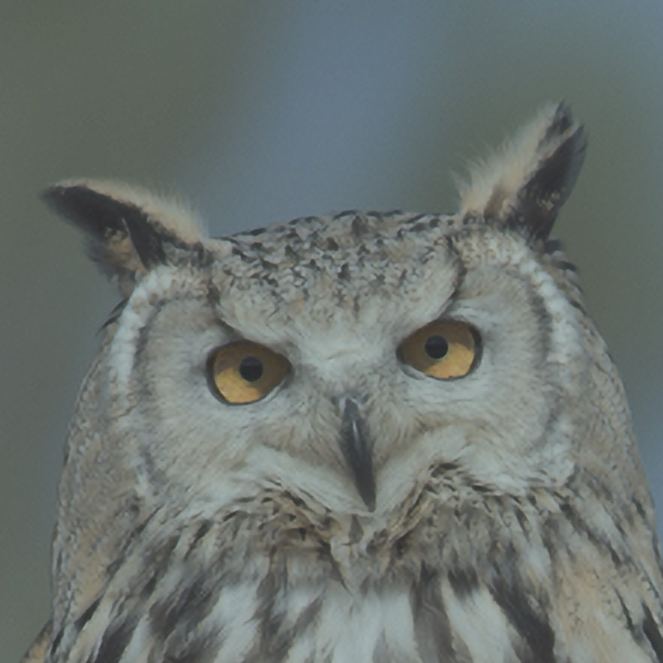"}}\\
\captionsetup[subfigure]{position=bottom}
\subfloat[(g) XLSR]{\includegraphics[width=0.17\linewidth]{"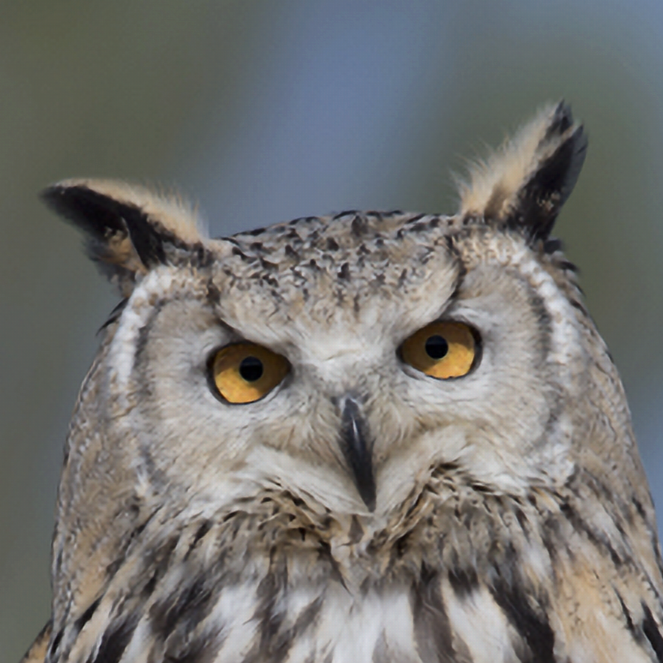"}}
\end{tabular}
\end{tabular}


\caption{Comparative results of UINT8 quantized models on DIV2KVal dataset. The proposed method is applied for (c)'s representative dataset, whereas all DIV2KVal images are used for (d) and (f)'s. (e) and (g) are the pre-trained quantized models provided by the authors. Visual results of (c), (e), and (g) are indistinguishable; however, XCAT runs faster}
\label{fig:div2k_photos}
\end{figure}

During the development of XCAT, many modified versions were created and tested. Numerical results of XCAT models and the ablation study done for HXBlocks are given in Tab.~\ref{tab:psnr}. Comparative visual results are given in Fig.~\ref{fig:div2k_photos}.

To choose the most successful model, we used the score function in (\ref{eq:customscore}), which is officially published in the competition's evaluation criteria.

\begin{equation}
\label{eq:customscore}
Score = \frac{2^{2(PSNR_{(UINT8)}-30)}}{t_{(UINT8)}10^{-5}}
\end{equation}

\textbf{Comparative study.} Tab.~\ref{tab:PSNR_drop} and Fig.~\ref{fig:div2k_photos} reveal that the network architecture's suitability for the quantization procedure plays a big role in producing high-quality, super-resolved images. Despite XLSR and ABPN having higher PSNR FP32 scores compared to XCAT, after the quantization, all three yielded similar visual results and closer UINT8 PSNR scores to each other.

\textbf{Ablation study.} In Tab.~\ref{tab:psnr}, increasing layer number/sizes and parameter numbers increased the PSNR score and decreased run time performance (E-G, G-L, B-C). Decreasing number of groups had a negative effect on PSNR; however, the positive effect on runtime surpassed (A-E). Using (I) dynamic kernels as opposed to not using (J) had a significant runtime boost with PSNR scores almost unchanged. Different heterogeneous divisions of filters (G-H) are also tried. Logically, when the input size of the 3x3 convolution layer increased, the PSNR score also increased. However, the penalty of runtime overcame the positive benefits of the PSNR raise. Replacing cross concatenation layers with 1x1 convolutions (XCAT-B) had the same effect as in the previous case.

Tab.~\ref{tab:XCrossMethods} shows the effect of using cross concatenation instead of directly concatenating the intermediate tensors in HXBlocks, as well as using different tensor divisions and number of HXBlocks. It is proven that using cross-concatenation allows for better information flow and increases the PSNR score, as opposed to using direct concatenation. This effect is more visible when the number of HXBlocks increase.  

\begin{table}
    \caption{PSNR (dB) drops for DIV2K(Val and Test, x3) before and after quantization, number of parameters, and runtime scores (ms) on Mali-G71 MP2 via AI Benchmark 5~\cite{AIBenchmark} and Synaptics NPU (*FP32 and UINT8 scores are taken from the paper. In addition, the authors' pre-trained .tflite model gave a concatenation error on AI Benchmark 4 \& 5, about the source tensor not being able to be used multiple times. Hence, the model code is altered for ABPN, where the relevant tensor is manually hard-copied and concatenated) ($^+$Tested in NCHW format)}
    \begin{center}
    \begin{tabular}{cccccc}
        \toprule
        \textbf{Metric} & \textbf{FSRCNN}~\cite{FSRCNN} & \textbf{ESPCN}~\cite{Depth2Space} & \textbf{XLSR}~\cite{XLSR} & \textbf{ABPN}~\cite{ABPN}* & \begin{tabular}{@{}c@{}} \textbf{XCAT} \\ (proposed) \end{tabular} \\
        \midrule
        {Val, FP32 PSNR} & {29.67} & {29.54} & {30.10} & {30.22} & \textbf{29.88} \\
        {Val, UINT8 PSNR} & {18.52} & {17.50} & {29.82} & {30.09} & \textbf{29.81} \\
        {$\Delta$PSNR} & {11.15} & {12.04 } & {0.28} & {0.13} & \textbf{\underline{0.07}}\\
        \midrule
        {Test, UINT8 PSNR} & {-} & {-} & {29.58} & {29.87} & \textbf{29.67} \\
        \midrule
        {\# of parameters} & {25K} & {31K} & {22K} & {42K} & \textbf{16K}\\
        {Synaptics Runtime$^+$} & {-} & {-} & {44.8} & {36.9} & \textbf{8.8}\\
        {Mali Runtime} & {485} & {363} & {370} & {600} & \textbf{320}\\
        {Score} & {0.003} & {0.061} & {210} & {188} & \textbf{240}\\ 
        \bottomrule
    \end{tabular}
    \end{center}
    \label{tab:PSNR_drop}
\end{table}

\section{Conclusions \& Future Studies}

This study proposes a lightweight, quantized single image super-resolution network named XCAT, submitted to \emph{Mobile AI \& AIM 2022 Real-Time Image Super-Resolution Challenge}. XCAT offers \textbf{heterogeneous group convolution blocks} which includes convolutional kernels with different kernels and input \& output tensor sizes. Compared to other studies which include group convolutions ending with 1x1 layers, \textbf{cross concatenating} the intermediate tensors between group convolutions offer runtime efficiency with tolerable sacrifice from PSNR test scores. To further increase runtime performance on mobile devices, upsampling done by tensor copy operations by default is replaced by a 1x1 convolutional layer with a non-trainable kernel. XCAT is also shown to be robust to quantization, with a decrease of 0.07dB from FP32 to the UINT8 model.

Comparative experimental results on slightly modified XCAT models reveal that the design choices proposed in this study offer the model to be deployed on mobile devices efficiently. To further prove the effectiveness of the proposed method, XCAT is evaluated with standardized datasets in comparison to other mobile-friendly super-resolution networks. Visual results indicate that XCAT can produce super-resolved images nearly identical to the other slower networks' outputs. Although HXBlock is designed for super-resolution problems, we believe that it can help many heavy models to facilitate running on mobile devices.

%
%
\bibliographystyle{splncs04}

\end{document}